\documentclass[10pt,letterpaper]{article}
\usepackage{opex3}

\newcommand{\ket}[1]{\mbox{$ | #1 \rangle $}}

\newcommand{\equationname}[1]{Eq.~#1}
\newcommand{\Sectionname}[1]{\textsc{Sec.~}#1}

\begin{document}

\title{Integrated optical source of polarization entangled photons at 1310 nm}

\author{A. Martin$^{1,\dagger}$, V. Cristofori$^{3,1}$, P.
Aboussouan$^1$, H. Herrmann$^2$,\\ W. Sohler$^2$, D.B. Ostrowsky$^1$,
O. Alibart$^1$, and S. Tanzilli$^{1}$}

\address{
$^1$Laboratoire de Physique de la Mati{\`e}re Condens{\'e}e, CNRS UMR 6622,\\
Universit{\'e} de Nice--Sophia Antipolis, Parc Valrose, 06108 Nice Cedex 2, France\\
$^2$Angewandte Physik, Universit\"at Paderborn, Warburger Str. 100,
33098 Paderborn, Germany\\
$^3$Dipartimento di Elettronica Informatica e Sistemistica, Universita di Bologna,\\
Viale Risorgimento 2, I 40136 Bologna, Italy }
 \email{$^{\dagger}$anthony.martin@unice.fr}

\begin{abstract}
We report the realization of a new polarization entangled
photon-pair source based on a titanium-indiffused waveguide
integrated on periodically poled lithium niobate pumped by a CW laser at $655\,nm$.
The paired photons are emitted at the telecom wavelength of $1310\,nm$ within
a bandwidth of $0.7\,nm$. The quantum properties of the pairs are
measured using a two-photon coalescence experiment showing a
visibility of 85\%. The evaluated source brightness, on the order
of $10^5$ pairs $s^{-1}\,GHz^{-1}\,mW^{-1}$, associated with its compactness and
reliability, demonstrates the source's high potential for
long-distance quantum communication.
\end{abstract}


\section{Introduction}

Quantum communication takes advantage of single quantum systems, such as photons, to carry the quantum analog of bits, usually
called qubits. Quantum information is encoded on the photon's quantum properties, such as polarization or time-bins of
emission~\cite{Weihs-Tittel_Photonic_01}. Selecting two orthogonal states spanning the Hilbert space, for instance $\ket{H}$
(horizontal) and $\ket{V}$ (vertical) when polarization is concerned, allows encoding the $\ket{0}$ and $\ket{1}$ values of
the qubit. Moreover quantum superposition makes it possible to create any state $\ket{\psi} = \alpha \ket{0} + e^{i \phi} \beta
\ket{1}$, provided the normalization rule $|{\alpha}|^2 + |{\beta}|^2 = 1$ is fulfilled.

Entanglement is a generalization of the superposition principle to multiparticle qubit systems. Pairs of polarization entangled
photons (or qubits) can be described by states of the form
\begin{equation}
\ket{\psi^\pm} = \frac{1}{\sqrt{2}}\left[\ket{H}_1 \ket{V}_2 \pm \ket{V}_1\ket{H}_2\right] \Leftrightarrow
\frac{1}{\sqrt{2}}\left[\ket{0}_1 \ket{1}_2 \pm \ket{1}_1\ket{0}_2\right],
\label{entangled}
\end{equation}
where the indices 1 and 2 label the two involved photons, respectively. The interesting property is that neither of the two
qubits has a definite value. But as soon as one of them is measured, the associated result being completely random, the state
of \equationname{(\ref{entangled})} indicates that the other is found to carry the opposite value. There is no classical analog to
this purely quantum feature~\cite{BCHSH_69}. This particularity of quantum physics is a resource for quantum
communication systems such as quantum key distribution~\cite{Gisin_QKD-Review_02},
quantum teleportation~\cite{Marcikic_tel_03}, and entanglement swapping~\cite{Halder_Ent_Indep_07}.

In today's quantum communication experiments, spontaneous parametric down-con\-ver\-sion (SPDC) in non-linear bulk crystals
is the common way to produce polarization entangled photons~\cite{Kwiat_NewPolarSource_95,Kwiat_Ultrabright_99}.
However, since such experiments are getting more and more complicated, they require sources of higher efficiency together
with narrower photon bandwidths~\cite{Halder_Ent_Indep_07,Halder_NJP_08}. In addition, as soon as long-distance quantum communication is
concerned, the paired photons have to be emitted within one of the telecom windows, i.e. around $1310\,nm$ or
$1550\,nm$~\cite{DeRied_Tele_QRelay_04}.

The aim of this work is to unite all of the above mentioned features in a single source based on a titanium (Ti) indiffused
periodically poled lithium niobate (PPLN) waveguide and a CW pump laser at $655\,nm$.
We report for the first time the efficient emission of narrowband polarization
entangled photons at $1310\,nm$, showing the highest quality of two-photon interference (coalescence) ever reported in a similar
configuration~\cite{Suhara_TypeII_PPLNW,Fujii_TypeII_RidgeW}. In the following, we will first describe the principle of the source.
Then, we will detail the characterizations leading to the validation of the emitted photon wavelength and associated
bandwidth, and to the estimation of the source brightness. Afterwards, we will move on to the interferometric setup designed
to evaluate the quality of the quantum properties of the emitted pairs. This experiment amounts to a typical Hong-Ou-Mandel interference involving two photons~\cite{HOM_Dip_87}. We will finally discuss the results taking into account an additional observable, i.e. energy-time entanglement, which depends on the phase-matching condition.

\section{Principle of the polarization entangled photon-pair source}

\begin{figure}[h]
\centering
\includegraphics[width=0.7\textwidth]{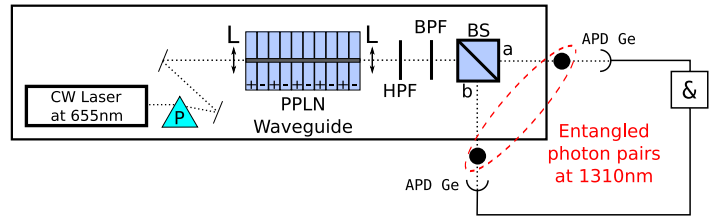}
\caption{Schematic of the polarization entangled photon-pair source at 1310\,$nm$. An external cavity diode laser at $655\,nm$ (Toptica Photonics DL100, $\Delta \nu \simeq MHz$, H-polarized) is employed to pump a Ti-indiffused PPLN waveguide in the CW regime; A prism (P) is used to remove the infrared light coming from the laser. A set of lenses (L) are used to couple light in and out of the waveguide.
The association of a high-pass filter (HPF, cut-off at $1000\,nm$, $T=90\%$) and a bandpass filter (BPF, $1310\,nm$, $\Delta \lambda = 10\,nm$, $T=70\%$) allows removing the residual pump photons. Finally, a 50/50 beam-splitter (BS) enables separating the paired photons, revealing entanglement in the coincidence basis. For characterization, we use two passively-quenched Ge-APDs connected to an AND-gate (\&) for coincidence counting.}
\label{Source}
\end{figure}

To date, the creation of entangled photon-pairs is usually performed by exploiting spontaneous parametric
down-conversion (SPDC) in non-linear bulk or waveguide crystals~\cite{Weihs-Tittel_Photonic_01}. The interaction of a
pump field ($p$) with a $\chi^{(2)}$ non-linear medium leads indeed, with a small probability, to the conversion of a pump
photon into so-called signal ($s$) and idler ($i$) photons. Naturally, this process is ruled by conservation of energy and
momentum
\begin{equation}
\left\lbrace
 \begin{array}{l c l}
  \omega_p &=& \omega_s + \omega_i\\
  \vec{k}_p &=& \vec{k}_s + \vec{k}_i + \frac{2\pi}{\Lambda}\cdot\vec{u},
 \end{array}
\right.
\end{equation}
where $\Lambda$ and $\vec{u}$ represent, in the specific case of any periodically poled crystal, the  poling
period and a unit vector perpendicular to the domain grating, respectively. Note that the latter equation is also known as
quasi-phase matching (QPM), which allows, compared to birefringent phase-matching in standard crystals, to compensate for dispersion
using the associated grating-type k-vector ($\frac{2\pi}{\Lambda}\cdot\vec{u}$). Then by an appropriate
choice of $\Lambda$, one can quasi-phase match practically any desired interaction within the transparency window of the
material. In this work, as depicted in \figurename{~\ref{Source}}, we take advantage of a Ti-indiffusion waveguide integrated on PPLN, for which the QPM condition has been chosen such that we expect, starting with a CW pump laser at $655\,nm$, the generation of pairs of photons at the telecom wavelength of $1310\,nm$. This way, for single photon counting, we can take advantage of passively-quenched Germanium
avalanche photodiodes (Ge-APDs) which do not require any additional gating signal on the contrary to
the experiments of~\cite{Suhara_TypeII_PPLNW,Fujii_TypeII_RidgeW}.

From the quantum side, since the generation of cross-polarized photons is necessary, the  waveguide device has to support both
vertical and horizontal polarization modes. Therefore, the well-established Ti-indiffusion technology can be applied for
waveguide fabrication and a type-II SPDC process, exploiting the $d_{24}$ non-linear coefficient of the material, can be
used~\cite{Suhara_TypeII_PPLNW}. Starting from H-polarized pump photons this process leads, at degeneracy, to the generation of
paired photons having strictly identical properties, but with orthogonal polarizations. As sketched in \figurename{~\ref{Source}}, after filtering out the remaining pump photons at the output of the crystal, the paired photons are directed to a 50/50 beam-splitter ($BS$) whose outputs are labelled $a$ and $b$. Such a device is used to separate the pairs, however four possibilities can occur at this stage. Two of them correspond to cases where the two photons exit through the same output port, $a$ or $b$, leading to states of the form $\ket{H}_a\ket{V}_a$ and $\ket{H}_b\ket{V}_b$, respectively. These two contributions are of no interest for our purpose. The two others correspond to cases where the two photons exit through different output ports, i.e. they are actually separated. In average, this separation occurs with a probability of $\frac{1}{2}$, but when successfull, the two possible output states, $\ket{H}_a\ket{V}_b$ and $\ket{V}_a\ket{H}_b$, have equal probabilities so that the related two-photon state corresponds to the entangled state of \equationname{(\ref{entangled})}. 
As SPDC ensures a simultaneous emission of the paired photons, the entangled state can be post-selected among all the other events using a coincidence detection scheme. Experimentally, such a coincidence basis can be defined by considering simultaneous detection events in modes $a$ and $b$ as depicted in \figurename{~\ref{Source}}. Three steps, i.e. SPDC, BS, and coincidence detection, are therefore cascaded for obtaining such a state configuration,
\begin{equation}
\begin{array}{c c l}
\ket{H}_p \stackrel{SPDC}{\,\,\,\,\,\longmapsto\,\,\,\,\,} \eta \ket{H}_s \ket{V}_i
&\stackrel{BS}{\,\,\,\,\,\longmapsto\,\,\,\,\,}&
\eta^{\ast} \frac{1}{2} \left [ \ket{H}_a \ket{V}_a + \ket{H}_b \ket{V}_b + \ket{H}_a \ket{V}_b + \ket{V}_a \ket{H}_b \right ]\\
&\stackrel{Coinc.}{\,\,\,\,\,\longmapsto\,\,\,\,\,}&
\eta^{\star} \frac{1}{\sqrt{2}} \left [ \ket{H}_a \ket{V}_b + \ket{V}_a \ket{H}_b \right ],
\label{2_steps}
\end{array}
\end{equation}
where $\eta$ and $\eta^{\ast}$ stand for the efficiencies of SPDC process and of the entire source, respectively. Note that $\eta^{\star}$
is obtained after renormalization.

\section{Fabrication of the PPLN waveguide, and characterization of the source\label{classic}}

To meet our goals, a 3.6$\,cm$ long sample with several Ti-indiffused waveguides of various widths (5, 6, and 7\,$\mu m$) was prepared. The waveguides were fabricated by an indiffusion of 104\,$nm$ thick Ti-stripes into a 0.5\,$mm$ thick Z-cut
LiNbO$_3$ substrate. Diffusion was performed at $1060^{\circ}C$ for 8.5~hrs. In this configuration, the required poling period for the generation of photon-pairs at the  degenerate wavelength of $1310\,nm$ was calculated to be around $6.6\,\mu m$. Subsequently, electric field-assisted periodic poling of the whole substrate was done with different periodicities (6.50 to 6.65\,$\mu m$ with steps of 0.05\,$\mu m$).
\begin{figure}[b]
\centering
\includegraphics[width=0.65\textwidth]{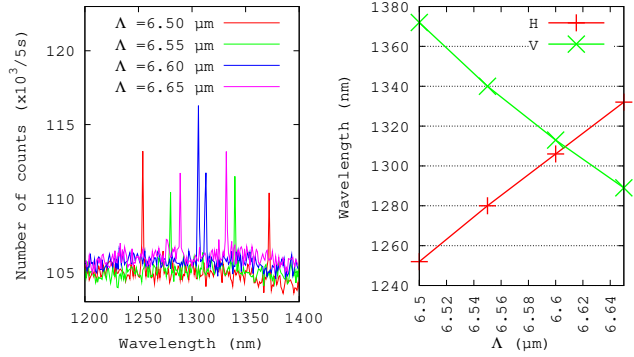}
\caption{Left: fluorescence spectra for various poling periods out of $7\,\mu m$-wide waveguides obtained in the single photon
counting regime. For all these curves, the temperature of the sample is $70^{\circ}C$ and the pump wavelength is $655\,nm$.
Right: QPM curve as a function of the poling period ranging from $6.50$ to $6.65\,\mu m$  with steps of 0.05\,$\mu m$. The straight line is a guide for the eye.}
\label{Degeneracy_poling}
\end{figure}

\begin{figure}[t]
\centering
\includegraphics[width=0.55\textwidth]{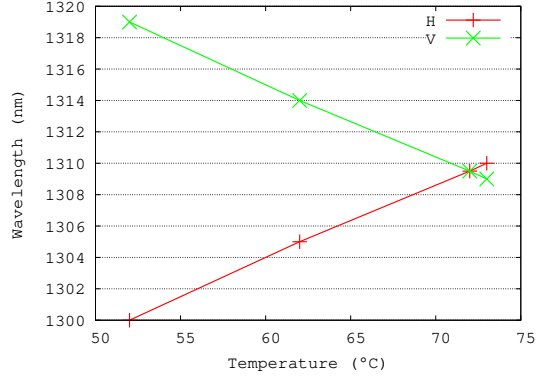}
\caption{QPM curve as function of the temperature for $\Lambda=6.60\,\mu m$. The degeneracy point can be reached by fine
tuning of the temperature up to $72^{\circ}C$. Note that before degeneracy, the longest wavelength is associated with the V
polarization mode, and the shortest to the H polarization, and vice-versa beyond degeneracy. The straight line is a guide for the eye.}
\label{Degeneracy_T}
\end{figure}

The first characterization of the sample concerns SPDC spectra that we measured in the single photon counting regime.
At an operating temperature of $70^{\circ}C$ and a pump wavelength of 655\,$nm$, photon-pair emission from 7\,$\mu$m
wide waveguides with different poling periodicities was observed as shown in \figurename{~\ref{Degeneracy_poling}}.
Following this, a fine tuning of the temperature up to $72^{\circ}C$ allowed us getting exactly both signal and idler photons at the degenerate wavelength of $1310\,nm$ out of a 6.6$\,\mu m$-period waveguide, as depicted in \figurename{~\ref{Degeneracy_T}}.

The measured bandwidth of those photons is very close to the resolution of our optical spectrum
analyzer ($0.6\,nm)$. After deconvolution with the monochromator resolution, we estimated the full-width at half-maximum (FWHM) bandwith to be approximately $0.7\,nm$. This result is in good agreement with the theoretical bandwidth calculated taking into account the $3.6\,cm$ length of our sample. Moreover, it has already been reported that type-II phase-matching~\cite{Suhara_TypeII_PPLNW,Fujii_TypeII_RidgeW} leads to much narrower bandwidths than type-0 or type-I phase-matching.
For instance, sources based on a proton-exchanged PPLN waveguide~\cite{Tanz_PPLW_01} and on a bulk $KNbO_3$ crystal~\cite{Tittel_10km_99} provided photon-pairs at 1310\,$nm$ within a bandwidth of 40 and 70\,$nm$, respectively. Our type-II PPLN waveguide therefore enables generating narrowband polarization photons. This is a clear advantage for long distance quantum communication since photons are less subject to both chromatic and polarization mode dispersions in optical fibers, preserving the purity of entanglement.

Another important figure of merit is the brightness of the source, i.e. the normalized rate at which the pairs are generated. The
commonly accepted brightness unit ($s^{-1}\,GHz^{-1}\,mW^{-1}$) is defined as the number of pairs produced per second, per $GHz$ of bandwith, and per $mW$ of pump power. Having a high-brightness source is of particular interest for both laboratory and practical quantum
communication experiments since low power, compact, and reliable pump diode lasers are sufficient for obtaining high counting
rates. Furthermore, if additional ultra-narrow filtering is necessary for some applications, having a very bright source still
enables using commercially available mid-power lasers as pumps~\cite{Halder_NJP_08}.
In the configuration of \figurename{~\ref{Source}}, the pair creation rate, $N$, has been estimated following the loss-independent method introduced in~\cite{Tanz_PPLW_01} where we find $N=\frac{S_aS_b}{2R_c}$.
Here $S_{a,b}$ and $R_c$ stand for the single and coincidence counting rates, respectively, when two single photon detectors are
placed after the BS in spatial modes $a$ and $b$. As already mentioned, we employ two passively-quenched Ge-APDs featuring 4\% detection efficiencies and 30\,$kHz$ of dark count rates. These APDs are connected to an AND-gate for coincidence counting.
Experimentally, we measured $S_a \approx S_b \approx 100 \cdot 10^3 \,s^{-1}$ and $R_c \approx 330\,s^{-1}$, and we estimated $N$
to be of about $1.5 \cdot 10^7 s^{-1}$. Then, taking into account a pump power of $0.4\,mW$ and a bandwidth of $0.7\,nm$, we
conclude our source emits $3 \cdot 10^5\,pairs\,\,\,s^{-1}\,GHz^{-1}\,mW^{-1}$. This high brightness result is mainly due to the waveguide configuration that permits confining the three waves, pump, signal, and idler, over longer distances than in bulk devices. Moreover, the reported brightness is of the same order as those reported in~\cite{Suhara_TypeII_PPLNW,Fujii_TypeII_RidgeW} for similar schemes.

\section{Quantum characterization of the source}

Obtaining polarization entangled photon-pairs (see \equationname{(\ref{2_steps})}) requires these two photons to be
indistinguishable for any degree of freedom, but the polarization, before they reach the BS of \figurename{~\ref{Source}}.
Especially, they have to arrive at the BS exactly at the same time with an accuracy better than their coherence time. However, since lithium niobate is birefringent, the two generated polarization modes do not travel at the same speed along their propagation in the waveguide. Knowing the length $L_{WG}$ and the birefringence of the waveguide, it is easy to calculate the average time delay between the two polarization modes :
\begin{equation} \left\langle \mathcal{T}_{delay}\right\rangle = \frac{L_{WG} \cdot \Delta n_{LiNbO_3}}{2c} \simeq 5\,ps,
\end{equation}
where $\Delta n _{LiNbO_3}$ corresponds to the difference of the group refractive indices for the two polarization
modes, and $c$ the speed of light. The spectrum analysis (see \figurename{~\ref{Degeneracy_poling}} and related discussion)
allows estimating a coherence time on the order of
\begin{equation} \mathcal{T}_{coh} = 0.44 \times
\frac{1}{c}\frac{\lambda^2}{\Delta\lambda} \simeq 3.6\,ps.
\label{t_coh}
\end{equation}
Comparing these two values indicates there is no temporal overlap at the beam-splitter for the generated photons, making them distinguishable. This leads to a separable two-photon state after the beam-splitter. In this context, a birefringent crystal placed between the waveguide and the beam-splitter would be necessary to compensate for the propagation time mismatch and to recover the desired entangled state at the output of the source. Without such a compensation crystal, a standard Bell test based on two polarization analyzers and a suitable coincidence detection apparatus cannot be employed to characterize entanglement~\cite{Kwiat_NewPolarSource_95,Suhara_TypeII_PPLNW}.
Provided these two photons are turned indistinguishable, it is nevertheless possible to infer the potential amount of entanglement by making them coalesce at a BS using a HOM type setup~\cite{HOM_Dip_87}. Indistinguishabi\-lity means that the photons have the same wavelength, bandwidth, polarization state, and spatial mode. In this case, if the two photons enter the BS through different inputs at the same time, the destructive interference makes them exit the device through the same output. Consequently, no coincidences are expected when two detectors are placed at the output of the BS.
\begin{figure}[h]
\centering
\includegraphics[width=0.70\textwidth]{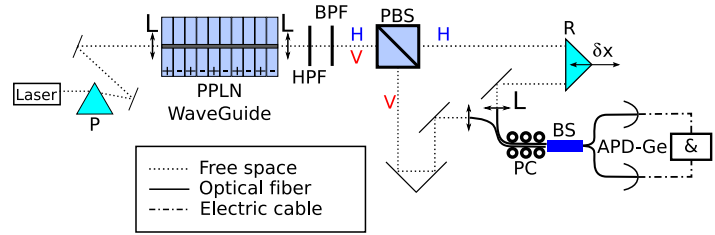}
\caption{Two-photon interference experiment. The two polarization modes are first separated using a polarization beam-splitter
(PBS). A retroreflector (R) placed in one arm is employed to adjust the relative delay of the two photons. After being coupled
into single mode optical fibers, these photons are recombined at a 50/50 coupler (BS) where quantum interference occurs. Note that
both polarization modes are adjusted to be identical using fiber-optics polarization controllers (PC) in front of the
coupler. The overall losses of the interferometer were estimated to be of 5.5\,$dB$.}
\label{MZ}
\end{figure}

Our interferometric apparatus, made of both free space and fiber-optics components, is depicted in \figurename{~\ref{MZ}}.
On the contrary to \figurename{~\ref{Source}}, a polarization beam splitter (PBS) is used to separate the paired photons
into two spatial modes regarding their polarization states (H,V). A motorized retroreflector and two polarization
controllers are employed to erase any temporal and polarization distinguishability before the two photons reach the 50/50 BS.
Two Ge-APDs connected to an AND-gate permit recording the single and coincidence rates as a function of the path length difference which is adjusted thanks to the retroreflector.
A so-called HOM dip in the coincidence rate ($R_C$) is expected when the arrival times of the photons at the BS are identical. Here, two parameters are of interest. On the one hand, the visibility (or depth), which depends on any experimental distinguishability, is the figure of merit which is linked to the quality of the entangled state produced by the source of \figurename{~\ref{Source}}. 
On the other hand, the width of the dip is directly related to the coherence time of the single photons~\cite{HOM_Dip_87}.

\begin{figure}[h]
\centering
\includegraphics[width=0.7\textwidth]{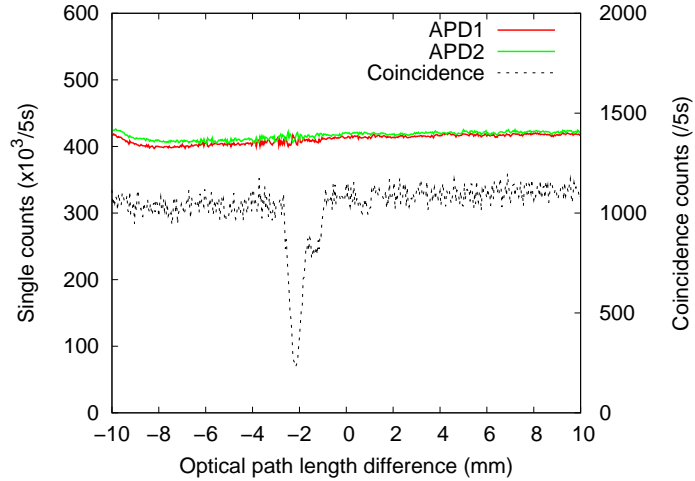}
\caption{Net coincidence and single counting rates at the output of the 50/50 beam-splitter as function of the relative length of the two arms.
Here the position of dip is linked to the relative separation experienced by the H and V photons in the generator due to their different group velocities. The dip exhibits a net visibility of $85\%$ and a width of $1.5\,mm$ FWHM for a temperature of $71.64^\circ C$.}
\label{dip}
\end{figure}

Figure~\ref{dip} exhibits the coincidence rate as a function of the path length difference between the two arms and
clearly shows a HOM interference while single photon detection remains constant in both APDs. The net
visibility can estimated from the coincidence curve following the relation
$V_{net} = \frac{R_C^{max}-R_C^{min}}{R_C^{max}-R_C^{acc}},$
where $R_C^{max}$ and $R_C^{min}$ are the coincidence rates outside and inside the dip, respectively. 
In our experiment, the accidental coincidence rate $R_C^{acc}$ is mostly due to having a dark count in both APDs simultaneously, and has been measured to be of about 100 per $5\,s$ integration time.
When the two photons characteristics are carefully adjusted to be identical, the net visibility is of about $85\%$.
To our knowledge, this result is the best ever reported for similar configurations, i.e.
waveguide-based sources emitting polarization entangled photons at telecom wavelengths~\cite{Suhara_TypeII_PPLNW,Fujii_TypeII_RidgeW}.
Moreover, when the distorsion of the dip is taken into account (see discussion in the next section), the full width at half maximum is estimated to be $1.5\,mm$. According to~\cite{HOM_Dip_87}, this corresponds to a coherence time of $3.5\,ps$ for the single photons which is in good agreement with the value previously obtained in \equationname{(\ref{t_coh})}.

\section{Discussion}

\begin{figure}[b] \centering
\includegraphics[width=0.75\textwidth]{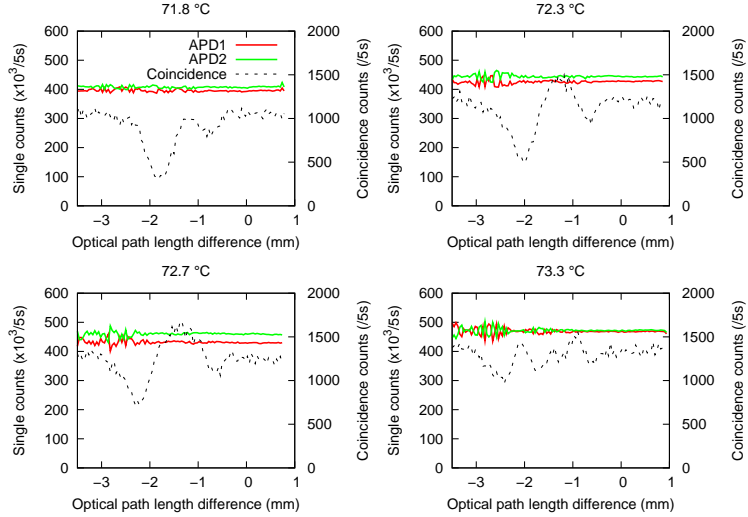}
\caption{Coincidence rate at the output of the 50/50 beam-splitter as  function of the relative length of the two arms for various
phase matching conditions leading to photons near degeneracy ($\Delta\lambda=\lambda_H-\lambda_V \leq 0.7\,nm$). It is then
interesting to note the decrease of the overall visibility, from $(a)$ to $(d)$, as the single photon wavelengths are tuned away
from degeneracy by an increase of the crystal temperature from $72$ to $73^{\circ}C$.}
\label{dip_away-dege}
\end{figure}

As we can see in \figurename{~\ref{dip}}, the HOM dip is noticeably  distorted and the bump on the right of the dip becomes
more important as we detuned the two photons central wavelength away from degeneracy. This effect is shown in
\figurename{~\ref{dip_away-dege}} where we observe an overall decreasing visibility as the two photons wave-functions less and
less overlap in terms of wavelength. Here we have to take into account another observable in the description of our two-photon
state to explain the observed beating in the dips of \figurename{~\ref{dip}} and \ref{dip_away-dege}. More precisely,
it is worth noting that CW SPDC naturally provides, in addition to any other observable, energy-time entangled photons.

It is known that two photons produced in the singlet Bell state $\ket{\psi^-}$, i.e. of  the form $\left[\ket{0}_1 \ket{1}_2
- \ket{1}_1\ket{0}_2\right]$, is the only state that gives rise to a coincidence peak when submitted to a HOM experiment due to
symmetry considerations. Such a state plays a crucial role in teleportation-like experiments where the BS acts as a Bell state
measurement apparatus~\cite{Marcikic_tel_03,Halder_Ent_Indep_07}.
Note that the three other Bell states that constitute the Bell basis lead to a HOM dip instead~\cite{BS_Zeilinger}. In our
source, the polarization modes of signal and idler photons are always associated with their wavelengths, as shown by the
quasi-phase matching curve of \figurename{~\ref{Degeneracy_T}}.
This no longer holds when the two photons are degenerate since they are produced in an entangled state of the form
\begin{equation}
\left|\frac{\omega_p}{2}+\delta\omega\right\rangle_H\left|\frac{\omega_p}{2}-\delta\omega\right\rangle_V+e^{\imath
2 \delta \omega \frac{\delta
x}{c}}\left|\frac{\omega_p}{2}-\delta\omega\right\rangle_H\left|\frac{\omega_p}{2}+\delta\omega\right\rangle_V
\label{state}
\end{equation}
where the frequency $\delta\omega$ is spanning over a frequency bandwidth corresponding to the $0.7\,nm$
obtained in \Sectionname{\ref{classic}}, and $\frac{\delta x}{c}$ the relative difference in arrival time between signal and idler
photons at the BS~\cite{dip_int_Kim}. It therefore comes, for a particular value $\delta x^-$, $e^{\imath 2 \delta \omega
\frac{\delta x^-}{c}}=-1$ making \equationname{(\ref{state})} become a $\ket{\psi^-}$ state which is not fully reachable in our
specific configuration. As a result, we believe the beating in the coincidence rate can be seen as the sum of a bump signature and a
dip signature as a function of $\delta x$ for a given $\delta \omega$. The overlapping part of our produced two-photon state
with $\ket{\psi^-}$ is responsible for the bump, the remaining non-overlapping part being responsible for the dip. We can
therefore conclude that the phase-matching condition in our waveguide gives rise to a state that partially overlaps with the
$\ket{\psi^-}$ state for the energy-time observable~\cite{dip_int_eckstein}. Figure~\ref{dip_away-dege} clearly indicates that tuning the emitted wavelengths does change the overlap of the actual created
two-photon state with $\ket{\psi^-}$. This means that the wavelengths of the photons have to be perfectly controlled to
avoid such an effect which is moreover not an issue in typical quantum communication based on polarization entangled qubits. In
any case, we have a clear signature that a high quality of polarization entanglement can be expected from the setup of
\figurename{~\ref{Source}}. Performing a Bell test experiment, together with a compensation crystal, would be a next step to properly
characterize the entanglement created by our source.

Finally note that Okamoto and co-workers reported the engineering of a partial $\ket{\psi^-}$ state at  degeneracy using high-order
phase dispersion in a bandpass filter placed on the path of one of the two photons~\cite{dip_int_okamoto}. They observed comparable
results to that of \figurename{~\ref{dip}}, i.e. an asymmetry in their HOM dip. In our case however, this possibility has to be excluded since both photons go through the bandpass filter, cancelling the dispersion effect.
Moreover, a run without the filter led to a similar shaped dip with a lower visibility due to higher background noise.

Work is currently in progress to address this interesting feature of the source.

\section{Conclusion and prospects}

Using a type-II PPLN waveguide, we have demonstrated a narrowband and bright source of  cross-polarized paired photons emitted at
1310\,$nm$ within a bandwidth of 0.7\,$nm$. We estimated the normalized production rate to be on the order of
$10^5\,pairs/s/GHz/mW$ which is one of the best ever reported for similar configurations~\cite{Suhara_TypeII_PPLNW,Fujii_TypeII_RidgeW}.
Furthermore, using a HOM-type setup, we obtained an anti-coincidence visibility of $85\%$ indicating a high level of
photon indistinguishability. To our knowledge, this visibility is the best ever reported for similar configurations.
These results, together with the compactness and reliability of the source, make it an high-quality generator of polarization entangled photon-pairs for the first time at $1310\,nm$.
This work clearly highlights the potential of integrated optics to serve as key elements for long-distance quantum communication protocols.

\section*{Acknowlegments}

This work was funded by the European ERA-SPOT program ``\textsc{Wasps}''.\\
V. Cristofori acknowledges ERASMUS program for her travelling and accommodation grant.\\
The authors thank P. Baldi and M.P. De Micheli for fruitful discussions.

\newpage

\end{document}